\documentclass[]{aa}  

\usepackage{graphicx}
\usepackage{txfonts}
\usepackage[]{hyperref}
\usepackage{caption}
\usepackage{subcaption}
\usepackage{gensymb}
\usepackage[dvipsnames]{xcolor}

\begin{document}

   \title{Planet-driven spirals in protoplanetary discs: limitations of the semi-analytical theory for observations}

   %\subtitle{I. Overviewing the $\kappa$-mechanism}

   \author{D. Fasano
          \inst{\ref{inst1},\ref{inst2}}
          \and
          A. J. Winter
          \inst{\ref{inst2}}
          \and
          M. Benisty
          \inst{\ref{inst1},\ref{inst2}}
          \and
          G. Rosotti
          \inst{\ref{inst3}}
          \and
          A. Ruzza
          \inst{\ref{inst3}}
          \and
          G. Lodato
          \inst{\ref{inst3}}
          C. Toci
          \inst{\ref{inst5},\ref{inst6}}
          \and
          T. Hilder
          \inst{\ref{inst4}}
          \and
          A. Izquierdo
          \inst{\ref{inst7}\ref{inst8}}
          \and
          D. Price
          \inst{\ref{inst4}}
          \fnmsep
          }

   \institute{Univ. Grenoble Alpes, CNRS, IPAG, 38000 Grenoble, France\\
              \email{daniele.fasano@oca.eu}\label{inst1}
        \and
            Université Côte d'Azur, Observatoire de la Côte d'Azur, CNRS, Laboratoire Lagrange, France\label{inst2}
        \and
            Dipartimento di Fisica, Università degli Studi di Milano, Via Celoria 16, Milano, Italy\label{inst3}
        \and
            School of Physics and Astronomy, Monash University, Clayton, Vic 3800, Australia\label{inst4}
        \and
            European Southern Observatory (ESO), Karl-Schwarzschild-Strasse 2, 
              85748 Garching bei Munchen, Germany\label{inst5}
        \and 
            INAF, Osservatorio Astrofisico di Arcetri, 50125 Firenze, Italy\label{inst6}  
        \and
           Leiden Observatory, Leiden University, P.O. Box 9513, 2300 RA Leiden, The Netherlands\label{inst7}
        \and
             Department of Astronomy, University of Florida, Gainesville, FL 32611, USA\label{inst8}
             }

   \date{Received March 22, 2024; accepted MM DD, 2024}

% \abstract{}{}{}{}{} 
% 5 {} token are mandatory
 
  \abstract
   {Detecting protoplanets during their formation stage is an important but elusive goal of modern astronomy. Kinematic detections via the spiral wakes in the gaseous disc are a promising avenue to achieve this goal.}
   {We aim to test the applicability to observations in the low and intermediate planet mass regimes of a commonly used semi-analytical model for planet induced spiral waves. In contrast with previous works which proposed to use the semi-analytical model to interpret observations, in this study we analyse for the first time both the structure of the velocity and density perturbations.}
   {We run a set of \textsc{FARGO3D} hydrodynamic simulations and compare them with the output of the semi-analytic model in the code \textsc{wakeflow}. We divide the disc in two regions: we use the density and velocity fields from the simulation in the linear region, where density waves are excited. In the nonlinear region, where density waves propagate through the disc, we then solve Burgers' equation to obtain the density field, from which we compute the velocity field.}
   {We find that the velocity field derived from the analytic theory is discontinuous at the interface between the linear and nonlinear regions. After $\sim 0.2~r_{\rm p}$ from the planet, the behaviour of the velocity field closely follows that of the density perturbations. In the low mass limit, the analytical model is in qualitative agreement with the simulations, although it underestimates the azimuthal width and the amplitude of the perturbations, predicting a stronger decay but a slower azimuthal advance of the shock fronts. In the intermediate regime, the discrepancy increases, resulting in a different pitch angle between the spirals of the simulations and the analytic model.}
   {The implementation of a fitting procedure based on the minimisation of intensity residuals is bound to fail due to the deviation in pitch angle between the analytic model and the simulations. In order to apply this model to observations, it needs to be revisited accounting also for higher planet masses.}


   \maketitle
%
%-------------------------------------------------------------------

\section{Introduction}

The Atacama Large (sub-)Millimiter Array (ALMA) has proven to be fundamental in the study of stars and planet formation. Images of the mm dust emission have shown the presence of many substructures (rings, gaps, cavities and spirals) in the vast majority of targeted discs \citep{Andrews_2018}. Although these features can be associated with the presence of unseen planets, planet-disc interaction is not the only explanation behind the formation of these substructures. For instance, rings and gaps can be due to magneto-rotational instabilities \citep{Flock_2015} or dust sintering outside the snow line \citep{Okuzumi_2012}. It is then difficult to unambiguously associate dust substructures with the presence of planets in disc and study their properties. However, ALMA opened yet another window to study protoplanetary discs by means of kinematic signatures. As the planet interacts with the gas, it excites spiral density waves and perturbs the disc velocity field leaving characteristic signatures in the molecular line emission \citep{Visual_Kin, Bollati_2021, Pinte_2023}.

Kinematic studies searching for planet-disc interaction signatures in molecular line emission have been carried out in recent years following different strategies. The effects of an embedded planet on the background velocity field may consist in deviations from Keplerian velocity in rotation curves of the gas \citep{Teague_2018}, a Doppler flip in the first moment map \citep{Perez_2018} and deviations from the isovelocity curves (so called kinks) in the channel maps \citep{Perez_2015}. This last technique was successfully used to infer the presence of protoplanets in the systems HD 163296 \citep{Pinte_2018} and HD 97048 \citep{Pinte_2019}, using the CO isotopologues line emission. At the moment, there are more than ten planet candidates waiting for confirmation observed with this method \citep{Pinte_2023}.

However, at the moment, the only way to obtain an estimate of the planet mass through this technique resides on the comparison between the observations and computationally expensive and time consuming numerical simulations \citep{Pinte_2019}. In the case of HD 163296 and HD 97048, planet mass estimates lie between 2 and $3~M_{\rm J}$. As this procedure relies on numerical hydrodynamical simulations, it is not possible to quantify the error on this measurements in a statistical manner due to the high computational cost. In order to develop a statistically significant procedure for the evaluation of the planet mass using MCMC (Markov Chain Monte Carlo) methods, an analytical model for planet produced kinks is needed.  

Planetary kinks are a result of the density waves perturbations generated by the planet. The first linear theory describing the propagation of density waves in a gaseous disc excited by a planet was developed by \citet{Goldreich_1979,Goldreich_1980} and \citet{Papaloizou_1984}. The perturbing planet excites density waves at Lindblad resonant locations, which undergo constructive interference and form the planetary spiral wake \citep{Ogilvie_2002}. \citet{Goodman_2001} and \citet{Rafikov_2002} first complemented the linear theory in a shearing box for the excitation of planetary spiral wakes with a nonlinear, semi-analytic framework describing its propagation. \citet{Miranda_2019, Miranda_2020} then generalised the linear theory, moving from the shearing sheet approximation to a global model. Finally, the nonlinear model was expanded in order to account also for the velocity perturbations in \citet{Bollati_2021}.

The linear excitation of the wake close to the planet and its nonlinear propagation in the rest of the disc can be naturally separated in two regimes when the planet mass satisfies $M_{\rm p}\la m_{\rm th}$, where the \textit{thermal mass} $ m_{\rm th}$ is:
\begin{equation}
    m_{\rm th} = \frac{2}{3}\frac{c_{\rm s}^3}{\Omega G} = \frac{2}{3}\left(\frac{h_{\rm p}}{r_{\rm p}}\right)^3M_\star,
    \label{eq:Thermal Mass}
\end{equation}
and $c_{\rm s}$ is the sound speed, $\Omega$ is the orbital frequency, $h_{\rm p}$ is the disc scale height, $r_{\rm p}$ is the planet radial position and $M_\star$ is the stellar mass. The thermal mass for realistic parameters ($h_{\rm p}/r_{\rm p}~\sim~0.1$, $M_\star\sim1~M_\odot$) is generally of the order of a Jupiter mass, which is in the range of the planet mass estimates obtained from the observations \citep{Pinte_2018,Pinte_2019}. The accuracy of the model in the low planet mass limit has been tested in \citet{Cimerman_2021}, focusing only on the density structure of the spiral.

However, observed planet candidates from continuum substructures \citep{Bae_2023,Lodato_2019} and molecular line emission \citep{Pinte_2023} need to be massive enough to create strong detectable signatures, often exceeding the thermal mass. With this paper, we aim to assess the applicability to observations of the model when approaching the thermal mass ($M_{\rm p}\sim1~m_{\rm th}$), a region of interest for currently ongoing observational surveys such as the ALMA large program exoALMA. In contrast with previous works which proposed to use the semi-analytical model to interpret observations, we focus on the velocity field in our analysis, as this is the quantity observed directly in the kinematic campaigns. In order to do so, we perform 2D numerical hydrodynamical simulations of a gaseous protoplanetary disc, considering different values of the planet mass, to be compared with the results obtained with the semi-analytical model. We quantitatively analyse the profiles of the azimuthal perturbations, focusing on the pitch angle of the spirals.

%Our future plan is to combine \textsc{wakeflow} with the state of the art for a systematic and statistically robust analysis of disc kinematics, the \textsc{discminer} package \citep{Izquierdo_2021, Izquierdo_2023}. \textsc{discminer} has proven to be an excellent and successful tool in the extraction of disc geometrical and line profile properties from kinematic observations. However, as the \textsc{discminer} model assumes only a Keplerian velocity field, it is not able to estimate the mass of the planet candidates. By adding the perturbation fields computed from \textsc{wakeflow}, it will then be possible to constrain also this parameter.

This paper is organised as follows: in Section~\ref{sec:Methods} we briefly summarise the semi-analytical model and we describe the numerical setup we used to test its accuracy. Section~\ref{sec:Results} presents the comparison between the model and the hydrodynamical simulations. Finally, we draw our conclusions in Section~\ref{sec:Conclusions}.

%--------------------------------------------------------------------
\section{Methods}
\label{sec:Methods} 

In this section we summarise the key points of the semi-analytic model, the main improvements with respect to previous iterations and the numerical setup of the hydrodynamic simulations.  

\subsection{Semi-analytical model}

\subsubsection{Background disc}
\label{sec:Background disc} 

We consider an unperturbed 2D disc in a cylindrical coordinate system $(r,\varphi)$, rotating around a star with mass $M_\star$. The hydrodynamical variables need to solve the following equations:
\begin{align}
    &\frac{\partial\Sigma_0}{\partial t} + \nabla \cdot (\Sigma_0\mathbf{v}) = 0,
    \label{eq:Continuity equation}\\
    &\frac{\partial\mathbf{v}}{\partial t} +(\mathbf{v}\cdot\nabla)\mathbf{v} = -\frac{1}{\Sigma_0}\nabla P - \nabla\Phi,
    \label{eq:Euler equation}
\end{align}
where $\Sigma_0$ is the surface density, $\mathbf{v}=(v_r,v_\varphi)$ the velocity field, $P$ the pressure of the gas and $\Phi\equiv -GM_\star/r$ is the gravitational potential of the star. 

We assume radial power laws for the sound speed $c_0$ and the surface density
\begin{align}
    &c_0 = c_{\rm p}\left(\frac{r}{r_{\rm p}}\right)^{-q},
    \label{eq:cs power law}\\
    &\Sigma_0 = \Sigma_{\rm p}\left(\frac{r}{r_{\rm p}}\right)^{-p},
    \label{eq:Sigma power law}
\end{align}
where $c_{\rm p}\equiv c_0(r_{\rm p})$ and $\Sigma_{\rm p} \equiv \Sigma_0(r_{\rm p})$. 

We consider a polytropic equation of state

\begin{equation}
    \label{eq: EDS poly}
    P = P_0\left(\frac{\Sigma}{\Sigma_0}\right)^\gamma
\end{equation}
with $\gamma$ the adiabatic index, that implies a perturbed sound speed

\begin{equation}
    \label{eq:perturbed sound speed}
    c^2 = \frac{\partial P}{\partial \Sigma} = c_0^2(r)\left(\frac{\Sigma}{\Sigma_0}\right)^{\gamma-1}.
\end{equation}
The model we will introduce in Sec.~\ref{sec:Semi-analytical model} is built assuming Eqs.~\ref{eq: EDS poly}-\ref{eq:perturbed sound speed}. We then take the limit for $\gamma\rightarrow1$ to obtain locally isothermal discs and fix $q=0$ to recover globally isothermal discs.

We define the disc aspect ratio 
\begin{equation}
    \frac{h}{r} \equiv \frac{c_0}{v_{\rm K}} = \frac{h_{\rm p}}{r_{\rm p}}\left(\frac{r}{r_{\rm p}}\right)^{1/2-q},
\end{equation}
where we used Eq.~\ref{eq:cs power law} in the second equality, $v_{\rm K} = (GM_\star/r)^{1/2}$ is the Keplerian velocity and we defined $h_{\rm p}/r_{\rm p}\equiv c_{\rm p}(GM_\star/r_{\rm p})^{-1/2}$.

Then, for a Keplerian power law disc the gas rotates around the star with sub-Keplerian velocity, with radial and azimuthal components \citep{Armitage_2015}
\begin{align}
    &v_{r,0} = 0,
    \label{eq:Background vr}\\
    &v_{\varphi,0} = v_{\rm K}\left[1 - a\left(\frac{h_{\rm p}}{r_{\rm p}}\right)^2\left(\frac{r}{r_{\rm p}}\right)^b\right]^{1/2},
    \label{eq:Background vphi}
\end{align}
where we have defined the coefficient $a\equiv p+q+3/2$, the exponent $b\equiv1-2q$ and the disc aspect ratio at the planet location $h_{\rm p}/r_{\rm p}\equiv h/r(r_{\rm p})$.

\subsubsection{Planet induced perturbations}
\label{sec:Semi-analytical model} 

We model the density and velocity perturbations induced by a planet embedded in a 2D gas disc starting from the semi-analytical method outlined in \citet{Bollati_2021}, based on the framework first introduced in \citet{Goodman_2001} and \citet{Rafikov_2002}. We expand on this model by removing some of the original approximations from \citet{Bollati_2021}. Here we summarise the main points of this method, the limitations of the original approach of \citet{Bollati_2021} and how we improve on it.

The shape of the wake generated by the planet in the linear approximation, resulting from constructive interference of density waves excited at Lindblad resonant locations \citep{Rafikov_2002,Ogilvie_2002}, is given by
\begin{equation}
    \varphi_{\rm wake}^{\rm linear} = \varphi_{\rm p} + \mathrm{sgn}(r - r_{\rm p})\int_{r_{\rm p}}^r\frac{\Omega(r') - \Omega_{\rm p}}{c_0(r')}\mathrm{d}r',
	\label{eq:Spiral Wake}
\end{equation}
where $(r_{\rm p}, \varphi_{\rm p})$ are the planet coordinates in a polar reference frame centered on the star location, $\Omega(r)$ and $\Omega_{\rm p}\equiv\Omega(r_{\rm p})$ are the disc and planet angular velocities, respectively, and $c_0(r)$ is the sound speed of the unperturbed disc. Assuming Keplerian rotation and a constant disc aspect ratio, substituting the power law prescription (Eq.~\eqref{eq:cs power law}), Eq.~\eqref{eq:Spiral Wake} becomes \citep{Rafikov_2002}
\begin{align}
    \varphi_{\rm wake}^{\rm linear} = \varphi_{\rm p} + \mathrm{sgn}(r - r_{\rm p})&\left(\frac{h_{\rm p}}{r_{\rm p}}\right)^{-1}\left[\frac{(r/r_{\rm p})^{q-1/2}}{q-1/2}+\right.\nonumber\\
    &-\left.\frac{(r/r_{\rm p})^{q+1}}{q+1}-\frac{3}{(2q-1)(q+1)}\right].
\end{align}
This linear prescription does not take into account nonlinear effects arising after the perturbations shock. \citet{Cimerman_2021} introduced a nonlinear correction to the spiral wake in the form
\begin{equation}
    \varphi_{\rm wake}^{\rm nonlinear} = \varphi_{\rm wake}^{\rm linear} + \mathrm{sgn}(r-r_{\rm p})\Delta\phi_0\frac{h_{\rm p}}{r_{\rm p}}\sqrt{t-t_0},
	\label{eq:Spiral Wake Nonlinear}
\end{equation}
where $\Delta\phi_0\simeq1$. This value is obtained from a fit on their simulations in the range $0.05-0.5~m_{\rm th}$.

The computation of the density perturbation is carried out in two distinct spatial regions: in an annulus centered on the planet position, where the density and velocity fields are obtained under the linear approximation, and further away from it, where the structure of the wake is calculated in the nonlinear regime \citep{Rafikov_2002}. We define the separation between these two regimes as follows: 

\begin{equation}
    \label{eq:linear edge}
    r_\pm = r_{\rm p} \pm 2l_{\rm p}
\end{equation}
with
\begin{equation}
    \label{eq:linear scale}
    l_{\rm p} = \frac{2}{3}h_{\rm p}.
\end{equation}
This value has been chosen so that the linear region is large enough to include Lindblad resonances exciting the spiral waves but limits nonlinear effects that start appearing after the spiral waves shock \citep{Goodman_2001}.\footnote{This criteria was obtained for a specific case. A detailed study of its dependence on disc parameters is still missing.}

\subsubsection{Linear region}

In the original model of \citet{Bollati_2021}, the density and velocity perturbations in the linear regime were obtained assuming the linear and shearing sheet approximations, following the procedure from \citet{Goodman_2001}. However, by definition of the shearing sheet approximation, this linear solution is valid only in a square box centered on the planet position with sides of the order of $l_{\rm p}$. As a result, we find that the azimuthal extent of the box is not large enough to fully capture the density profile needed as an initial condition to solve Burgers' equation in the nonlinear regime.

%With these assumptions, it is possible to linearise Eqs.~\ref{eq:Continuity equation}-\ref{eq:Euler equation} and solve them in a local Cartesian coordinate system, neglecting the radial dependence of the physical quantities. In this way, one would obtain a solution that depends only on the planet mass and can be scaled linearly \citep{Goodman_2001,Bollati_2021}. 

An alternative approach has been presented in \citet{Miranda_2019, Miranda_2020}. In this framework the authors provide linear global density and velocity fields by solving a master equation for the enthalpy of the disc. In this way it is then possible to extract the density profile along the entire azimuthal range and use it as the initial condition for the computation of the density field in the nonlinear regime.

A third option consists in using both the density and velocity fields retrieved from numerical hydrodynamic simulations. \citet{Cimerman_2021} showed that the global linear solution is in good agreement with the simulation in the linear region and used the profile from the simulation as an initial condition to solve Burgers' equation. In this paper we follow the same approach and use the density and velocity fields from our hydrodynamical simulations (see Section~\ref{sec:Numerical setup}) in the annular region of width $4l_{\rm p}$ centered on the planet. 

%In the linear regime, the density and velocity perturbations are computed in a square with side $8h_{\rm p}/3$ centered on the planet position. In this region, the fluid equations are solved in local Cartesian coordinates under the shearing sheet approximation, taking the sound speed and surface density to be constant and equal to their value at the planet location, and the linear approximation, assuming $M_{\rm p}\ll m_{\rm th}$. These equations were solved in \citet{Bollati_2021} following the approach outlined in \citet{Goodman_2001}. As the perturbations are linear in the planet mass, we simply scaled the results by the parameter $M_{\rm p}/m_{\rm th}$.

\subsubsection{Nonlinear region}
In the nonlinear regime, the equations are solved in a polar reference frame co-rotating with the planet. \citet{Rafikov_2002} showed that the structure of the density perturbation outside the linear region is described by the inviscid Burgers' equation in a specific reference system with coordinates $t, \eta$
\begin{equation}
    \partial_t\chi + \mathrm{sgn}(r - r_{\rm p})\chi\partial_\eta\chi = 0,
    \label{eq:Burgers Equation}
\end{equation}
where
\begin{align}
    &t \equiv -\frac{r_{\rm p}}{l_{\rm p}}\frac{m_{\rm p}}{m_{\rm th}}\int_{r_{\rm p}}^r\frac{\Omega(r') - \Omega_{\rm p}}{c_0(r')g(r')}\mathrm{d}{r'},
    \label{eq:t}\\
    &\eta \equiv \frac{r_{\rm p}}{l_{\rm p}}[\varphi - \varphi_{\rm wake}(r)],
    \label{eq:eta}\\
    &\chi \equiv \frac{\gamma+1}{2}\frac{\Sigma-\Sigma_0}{\Sigma_0}g(r),
    \label{eq:Chi}\\
    &g(r) \equiv \frac{2^{1/4}}{r_{\rm p}c_{\rm p}\Sigma_{\rm p}^{1/2}}\left(\frac{r\Sigma_0c_0^3}{|\Omega-\Omega_{\rm p}|}\right)^{1/2},
    \label{eq:g}
\end{align}
with $\gamma$ the adiabatic index, $\chi$ represents the density perturbation and the $(t,\eta)$ coordinates represent the distance along the spiral wake and in the azimuthal direction with respect to the centre of the wake, respectively. In the case of a power-law Keplerian disc, using Eqs.~\eqref{eq:cs power law}-\eqref{eq:Sigma power law}, we can write Eqs.~\eqref{eq:t} and \eqref{eq:g} as
\begin{align}
    &t \equiv - \frac{3}{2^{5/4}(h_{\rm p}/r_{\rm p})^{5/2}}\frac{m_{\rm p}}{m_{\rm th}}\int_1^{r/r_{\rm p}}(1-x^{3/2})|1-x^{3/2}|^{1/2}x^w\mathrm{d}x,
    \label{eq:t power law}\\
    &g(r) \equiv 2^{1/4}\left(\frac{h_{\rm p}}{r_{\rm p}}\right)^{1/2}\frac{(r/r_{\rm p})^{5/4-(p+3q)/2}}{|1-(r/r_{\rm p})^{3/2}|^{1/2}},
    \label{eq:g power law}
\end{align}
where $w = -11/4 + (p+5q)/2$.

The radial and azimuthal velocity perturbations can be related to the density perturbation $\chi$ with the following expressions 
% \begin{align}
%     &v_r(r,\varphi) = \mathrm{sgn}(r - r_{\rm p})\Lambda_{\rm p}f_{v_r}(r)\chi(t,\eta),
%     \label{eq: u nonlinear}\\
%     &v_\varphi(r,\varphi) = \mathrm{sgn}(r - r_{\rm p})\Lambda_{\rm p}f_{v_\varphi}(r)\chi(t,\eta),
%     \label{eq: v nonlinear}
% \end{align}
% where, assuming a power-law Keplerian disc described by Eqs.~\eqref{eq:cs power law}-\eqref{eq:Sigma power law}, we have
% \begin{align}
%     &\Lambda_{\rm p}f_{v_r}(r) = c_{\rm p}\left(\frac{h_{\rm p}}{r_{\rm p}}\right)^{-1/2}\frac{2^{3/4}}{\gamma + 1}\left(\frac{r}{r_{\rm p}}\right)^{(p + q -1)/2}\left|\left(\frac{r}{r_{\rm p}}\right)^{-3/2}-1\right|^{1/2},
%     \label{eq:Lambda f u}\\
%     &\Lambda_{\rm p}f_{v_\varphi}(r) = c_{\rm p}\left(\frac{h_{\rm p}}{r_{\rm p}}\right)^{1/2}\frac{2^{3/4}}{\gamma + 1}\left(\frac{r}{r_{\rm p}}\right)^{(p - q -3)/2}\left|\left(\frac{r}{r_{\rm p}}\right)^{-3/2}-1\right|^{-1/2}.
%     \label{eq:Lambda f v}
% \end{align}

\begin{align}
    &v_r(r,\varphi) = \mathrm{sgn}(r - r_{\rm p})\frac{2c_0}{\gamma+1}\psi
    \label{eq: u nonlinear}\\
    &v_\varphi(r,\varphi) = \frac{c_0}{(\Omega-\Omega_{\rm p})r}v_r,
    \label{eq: v nonlinear}
\end{align}
where we defined the mathematical quantity
\begin{equation}
\label{eq:psi old}
    \psi \equiv \frac{\gamma+1}{\gamma-1}\frac{c-c_0}{c_0}.
\end{equation}
By substituting Eq.~\ref{eq:perturbed sound speed} we can express $\psi$ as a function of $\Sigma$
\begin{equation}
\label{eq:psi}
    \psi = \frac{\gamma+1}{\gamma-1}\left[\left(\frac{\Sigma}{\Sigma_0}\right)^{(\gamma-1)/2}-1\right].
\end{equation}

In the original model of \citet{Bollati_2021}, they assumed that the sound speed perturbation is a fraction of the unperturbed sound speed ($\psi\ll1$), and expanded to first order Eq.~\ref{eq:psi}, obtaining approximate formulas for Eqs.~\ref{eq: u nonlinear}-\ref{eq: v nonlinear}. We found that the condition $\psi\ll1$ is not always satisfied, so we consider the exact transformation Eq.~\ref{eq:psi} to compute the velocity field in this paper.

We solved Eq.~\eqref{eq:Burgers Equation} following the method outlined in \citet{Bollati_2021}, using the density perturbation profile evaluated at the edge of the linear annulus as an initial condition. However, \citet{Bollati_2021} solved Burgers' equation only in the outer disc (i.e. for $r>r_{\rm p}$) and then copied the result in the inner disc (i.e. for $r<r_{\rm p}$). We improved on their approach by solving the Burgers' equation both in the outer and the inner disc, using as an initial condition the density perturbation profile evaluated at the outer and inner edge of the linear annulus, respectively. The model was implemented in an open source code, \textsc{wakeflow}\footnote{https://github.com/DanieleFasano/wakeflow} \citep{Hilder_2023}, that will be used in the following. 

\subsection{Numerical setup}
\label{sec:Numerical setup} 

\begin{table}
\caption{Simulation Disc Parameters}             % title of Table
\label{table:Simulation params}      % is used to refer this table in the text
\centering                          % used for centering table
\begin{tabular}{c c c c c c c c}        % centered columns (4 columns)
\hline\hline                 % inserts double horizontal lines
Model & $M_{\rm p}$ & $M_{\rm p}$ & $h_{\rm p}/r_{\rm p}$ & $\alpha$ & $q$ & $p$ & $r_{\rm p}$\\    % table heading 
& $[M_{\rm J}]$ & $[m_{\rm th}]$ & & & & & [au] \\
\hline                        % inserts single horizontal line
   1 & 0.0218 & 0.25 & 0.05 & 1e-4 & 0 & 1.5 & 100\\      % inserting body of the table
   2 & 0.0872 & 1.00 & 0.05 & 1e-4 & 0 & 1.5 & 100\\
   3 & 0.0218 & 0.25 & 0.05 & 1e-3 & 0 & 1.5 & 100\\      % inserting body of the table
   4 & 0.0872 & 1.00 & 0.05 & 1e-3 & 0 & 1.5 & 100\\
   5 & 0.667  & 1.00 & 0.1  & 1e-3 & 0 & 1.5 & 100\\

\hline                                   %inserts single line
\end{tabular}
\end{table}

We perform 2D hydrodynamical simulations using the code \textsc{FARGO3D} \citep{Benitez_2016, Masset_2000}. We use the same setup described in \citet{Cimerman_2021} in order to compare our results in the low planet mass limit. Simulations are run in cylindrical coordinates on a meshed domain that extends radially between $0.2r_p$ and $4r_p$ with logarithmic spacing, and uniformly over the full disc azimuthal extension. The grid resolution is $N_\phi \times N_r = 14400\times 6896$, corresponding to a minimum of 50 cells per scale height at all simulated radii for each of our setups. We initialize each simulation in an axisymmetric state. To achieve centrifugal balance, we set the azimuthal component of the gas initial velocity to the Keplerian solution with the pressure support correction while we set to zero the radial component. We initialize the surface density prescribing the radial profile as a power law of index $-p$. We used closed boundary conditions implementing wave-killing zones near the radial boundaries for $0.2\leq r/r_p \leq 0.28$ and $3.4\leq r/r_p \leq 4$ to avoid reflections. In these regions, the gas density and radial velocity are relaxed towards their initial conditions, analogously to \citet{DeValBorro_2006}, on a timescale $\tau=0.3/\Omega(r)$.  

 We consider five simulations of a gas disc rotating around a $1~M_\odot$ star, hosting a non-migrating and non-accreting planet with mass $\{0.25, 1.0\}~m_{\rm th}$ to study the behaviour of the semi-analytic model in the low and intermediate mass regimes, respectively. To avoid spurious shocks, the planet mass is gradually increased up to its target value over ten orbits in the same way as in the work of \citet{Cimerman_2021}. The planet potential ($\Phi_p$) is smoothed near the planet position as $\Phi_p = -GM_p/\sqrt{r^2+s^2}$ with smoothing length $s=0.6\times H(r)$. We included the indirect term that arises from the acceleration of our non-inertial frame of reference centred on the star. We set the gas viscosity adopting the \citet{Shakura_Sunyaev_1973} \mbox{$\alpha$-prescription} with constant $\alpha = \{10^{-4}, 10^{-3}\}$. We simulated disc aspect ratios $h_{\rm p}/r_{\rm p} = \{0.05, 0.1\}$, fixed the surface density power law index $p = 1.5$ and used a globally isothermal equation of state to compare directly with the fiducial simulation in \citet{Cimerman_2021}. We summarise the parameters we used in our simulations in Table~\ref{table:Simulation params}.
We evolved the simulated discs for 100 orbits of the planet.

For each simulation, we use the semi-analytical model method outlined in Section~\ref{sec:Semi-analytical model} to calculate the corresponding density and velocity perturbations. We use the same disc parameters and planet position of the simulations, fixing $q=0$ and $\gamma = 1$ to reproduce the globally isothermal model. As $\psi$ (Eq.~\ref{eq:psi}) is not defined for $\gamma = 1$, we check that our results are not strongly dependent on the isothermal limit considering values of $\gamma=\{1.01, 1.001, 1.0001\}$, observing no changes in our results. We compute the solution on a Cartesian grid of $1024\times1024$ points.

%-----------------------------------------------------------------
\section{Results}
\label{sec:Results} 

%and high mass ($M_{\rm p}>m_{\rm th}$) 
In this paper we aim to numerically test the semi-analytical theory of planet induced spiral density waves. \citet{Cimerman_2021} previously tested the behaviour of the solution of Burgers' equation for a planet mass smaller than the thermal mass. They considered a globally isothermal disc with disc aspect ratio $h_{\rm p}/r_{\rm p} = 0.05$ hosting a planet with mass $M_{\rm p} = 0.25~m_{\rm th}\simeq 0.02~M_{\rm J}$. In this regime, they found that the solution of Burgers' equation is in qualitative agreement with their simulations, but it underestimates the amplitude of the density perturbations and the azimuthal propagation of the shock fronts. They also showed how the analytical model does not capture the presence of additional spiral arms in the inner disc. Secondary and tertiary spiral arms are a result of constructive interference of perturbations with different azimuthal modes, occurring outside of the linear region, thus they are not captured in the profile used as an initial condition for Burgers’ equation and cannot be propagated by the analytical model.

However, it is very difficult to observe planets in this regime as they excite low amplitude density and velocity spiral perturbations. For this reason, we expand on the study by \citet{Cimerman_2021} by considering also the intermediate ($M_{\rm p}\simeq m_{\rm th}$) limit. Moreover, kinematic observations of protoplanetary discs give direct constraint on the velocity field, not the density field of the gas. Thus, we consider the velocity field for the first time in our analysis. We also focus only on the outer disc, as the presence of additional spiral arms is not reproduced by the analytic model. Additionally, it is easier to resolve spiral perturbations in the outer disc from the observations. We also focus only on the radial velocity field. The azimuthal velocity component is one order of magnitude lower in amplitude, so the velocity field is dominated by the radial component.

%In Sec.~\ref{sec:Numerical setup} we described the setup we use for our numerical simulations. The choice of these parameters is driven mainly to perform a direct comparison with the analysis of \citet{Cimerman_2021}.

\subsection{Low mass limit}

   \begin{figure}
   \centering
       \begin{subfigure}[t]{\hsize}
       \centering
       \includegraphics[width=\hsize]{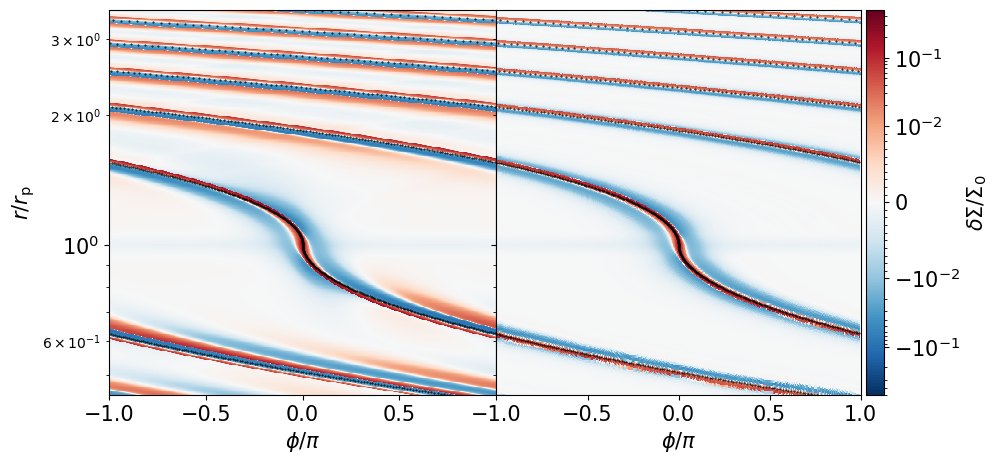}
          \caption{
                  }
             \label{fig:Density Map 0.25}
        \end{subfigure}
   \hfill 
       \begin{subfigure}[t]{\hsize}
       \centering
       \includegraphics[width=\hsize]{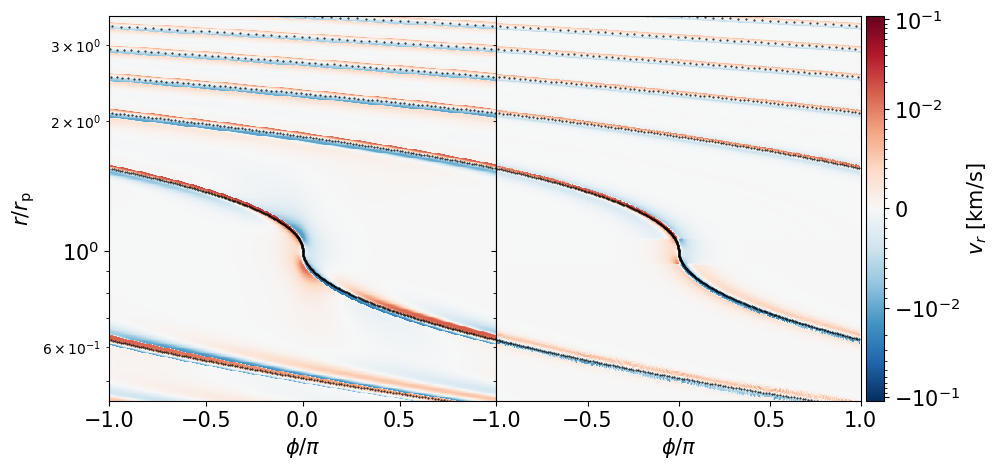}
          \caption{
                  }
             \label{fig:Velocity Map 0.25}
        \end{subfigure}
    \caption{$r$-$\phi$ maps of the density (a) and radial velocity (b) perturbation fields from the simulation (left panel) and analytical model (right panel) in the $M_{\rm p}=0.25\ m_{\rm th}$ case. The black dotted line represents the linear wake (Eq.~\ref{eq:Spiral Wake}). The colorbar is logarithmic above 0.01 km/s and linear below it. For reference, the Keplerian speed at the planet location is  $v_{\rm K}=3~\rm km/s$.
                  }
   \end{figure}

   \begin{figure*}
   \centering
   \resizebox{\hsize}{!}{
        \begin{subfigure}[t]{\textwidth}
        \centering
        \includegraphics[width=\textwidth]{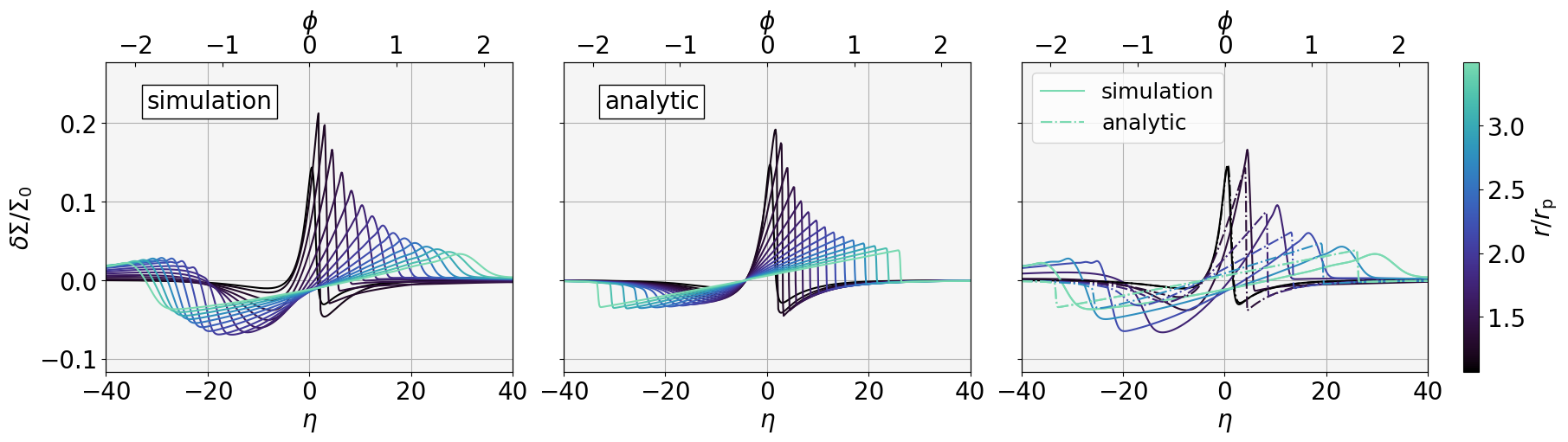}
        \caption{
              }
         \label{fig:Density Profiles 0.25}
        \end{subfigure}
        }
    \resizebox{\hsize}{!}{
        \hfill 
        \begin{subfigure}[t]{\textwidth}
        \centering
        \includegraphics[width=\textwidth]{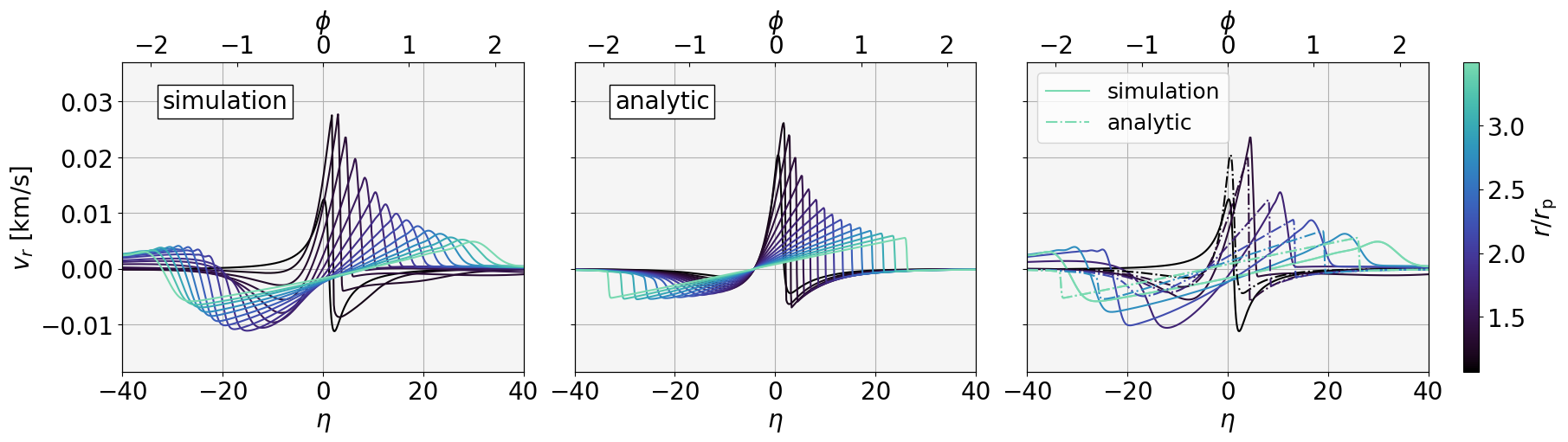}
        \caption{
              }
         \label{fig:Velocity Profiles 0.25}
         \end{subfigure}
        }
    \caption{Comparison between simulated (left) and analytical (middle) density (a) and radial velocity (b) $\eta$ (azimuthal) profiles in the outer disc for the $M_{\rm p}=0.25\ m_{\rm th}$ case. In the right panels we show a choice of profiles from the simulation (solid line) and the analytic model (dash dotted line) to better highlight the differences in individual profiles. The color scale shows the radial distance from the planet. We filter the profiles from the simulation using the savgol\_filter function from the scipy Python library to reduce the oscillations at the shock fronts (see Appendix~\ref{App}).
              }
   \end{figure*}

Here we present the results we obtain in the low planet mass limit ($m_{\rm p}\ll m_{\rm th}$). In Fig.~\ref{fig:Density Map 0.25} we compare the density perturbations extracted from the simulation against those computed using \textsc{wakeflow}. The black dotted curve represents the spiral wake given by the linear approximation (Eq.~\ref{eq:Spiral Wake}). In this regime the pitch angle of the spirals is comparable, however their amplitude and width are smaller in the analytic model. Moreover the simulation shows the presence of an additional positive spiral arm below the black dotted line.

We show the same comparison for the radial velocity field in Fig.~\ref{fig:Velocity Map 0.25}. The width and amplitude differences are less severe compared to the density perturbation. The analytic model features a discontinuity between the linear and nonlinear regions. This is because both the density and radial velocity in the linear regime are taken from the simulation, while only the density is obtained solving Burgers' equation in the nonlinear regime. The velocity components are then computed using Eqs.~\ref{eq: u nonlinear}-\ref{eq: v nonlinear}, which are obtained from Eqs.~\ref{eq:Continuity equation}-\ref{eq:Euler equation} neglecting higher order terms \citep{Rafikov_2002}. We believe  this simplification in the model to be the source of the discontinuity.

We compare the properties of the spiral more quantitatively by looking at the azimuthal ($\eta$) profiles of the density and radial velocity perturbations, shown in Fig.~\ref{fig:Density Profiles 0.25} and \ref{fig:Velocity Profiles 0.25}, respectively. The density profiles exhibits similar features to the ones presented in \citet{Cimerman_2021}. When compared with the simulation, the amplitude of the spiral computed from Burgers' equation suffers a stronger decay and the shocks advance more slowly in the azimuthal direction. Moreover, the analytic profiles feature steep shock fronts, in contrast with the smoother profiles from the simulation, which are due to diffusive effects introduced by viscosity. Viscous diffusion affects the perturbations by lowering their amplitude, while increasing their azimuthal extent. Indeed, the simulated profiles are wider than the analytic ones, but they still feature a higher amplitude. This suggests that the analytic model tends to overestimate the wave damping, in agreement with the results found in \citet{Cimerman_2021}.

While we note that using a simulation with very low viscosity ($\alpha\ll10^{-4}$) would provide an ideal comparison with the inviscid analytic model, we have noticed that setting such a low $\alpha$ parameter triggers the onset of numerical effects, requiring an even higher resolution in order to remove them. Although increasing viscosity up to $\alpha=10^{-3}$ removes these numerical artifacts, it also causes strong diffusion and makes a direct comparison between the simulated and analytic profiles more complicated. We find that choosing $\alpha=10^{-4}$ gives the best trade off between suppressing numerical effects and limiting viscous spreading. However, we still see the presence of small oscillations at the shock fronts. In Appendix~\ref{App} we use the $\alpha=10^{-3}$ simulations to check that these oscillations are a result of the low viscosity and do not affect our results.

%We note the presence of small oscillations at the shock fronts in the low viscosity ($\alpha=10^{-4}$) simulation profiles. In Appendix~\ref{App} we use the $\alpha=10^{-3}$ simulations to check that these oscillations are a result of the low viscosity and do not affect our results. However, a viscosity of $\alpha=10^{-3}$ is too high, causing strong diffusion and making a direct comparison between the simulated and analytic profiles more complicated. Although using a simulation with vanishing viscosity ($\alpha=10^{-6}$) would provide an ideal comparison with the inviscid analytic model, we have noticed that setting a lower $\alpha$ parameter triggers the onset of numerical effects, requiring an even higher resolution in order to remove them. For this reason, a viscosity of $\alpha=10^{-4}$ is the best trade off to carry out our analysis.

Additionally, we observe the presence of a positive peak for negative $\eta$ in the simulation. This affects both the amplitude and width of the profile, causing the strongest deviation between the simulation and the model. Although the velocity profiles are remarkably different close to the planet due to the discontinuity previously discussed, after about $0.2~r_{\rm p}$ their behaviour becomes very similar to that of the density perturbations. 

Lastly, we can define the center of the spiral as the point where the perturbations change sign, in the limit when the wavefronts have already undergone a shock and the spiral features its characteristic N shape. This point is fixed by construction in the analytic model. We note that this is not the case in the simulation, where it features a positive azimuthal shift. This implies that the pitch angle of the spiral from the analytical model is different with respect to that of the simulation at the center of the spiral.

\subsection{Intermediate mass limit}

   \begin{figure}
   \centering
        \begin{subfigure}[t]{\hsize}
        \centering
        \includegraphics[width=\hsize]{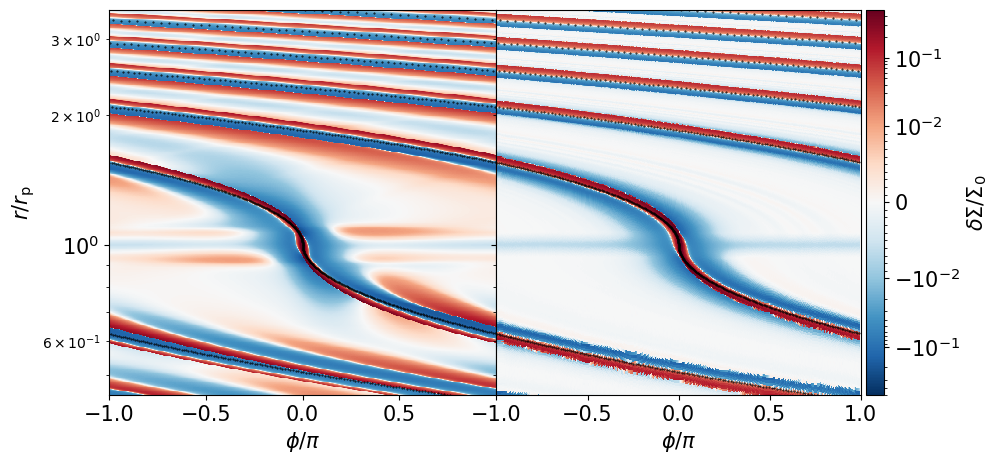}
        \caption{
              }
         \label{fig:Density Map 1.00}
        \end{subfigure}
   \hfill 
        \begin{subfigure}[t]{\hsize}
        \centering
        \includegraphics[width=\hsize]{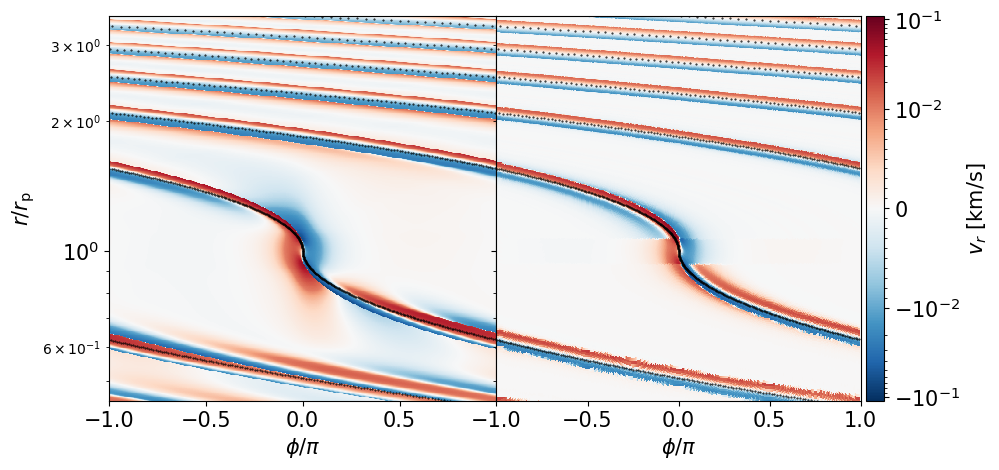}
        \caption{
              }
         \label{fig:Velocity Map 1.00}
         \end{subfigure}
    \caption{$r$-$\phi$ map of the density (a) and radial velocity (b) perturbation fields from the simulation (left panel) and analytical model (right panel) in the $M_{\rm p}=1.00\ m_{\rm th}$ case. The black dotted line represents the linear wake (Eq.~\ref{eq:Spiral Wake}). The colorbar is logarithmic above 0.01 km/s and linear below it. For reference, the Keplerian speed at the planet location is  $v_{\rm K}=3~\rm km/s$.
              }
   \end{figure}

   \begin{figure*}
   \centering
   \resizebox{\hsize}{!}{
        \begin{subfigure}[t]{\textwidth}
        \centering
        \includegraphics[width=\hsize]{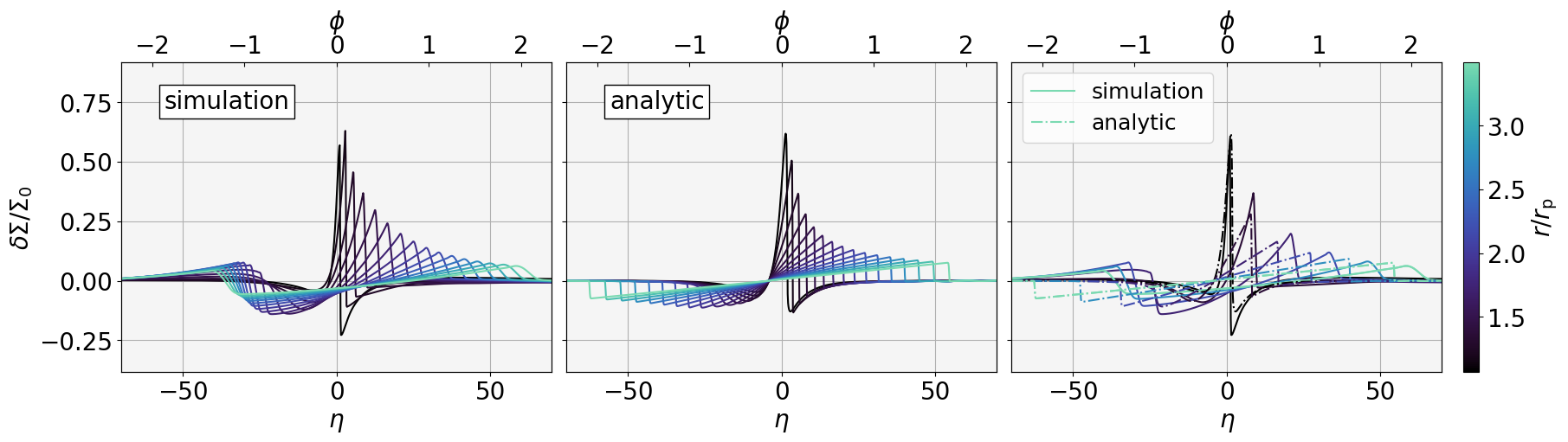}
        \caption{
              }
         \label{fig:Density Profiles 1.00}
        \end{subfigure}
        }
        \resizebox{\hsize}{!}{
        \hfill 
        \begin{subfigure}[t]{\textwidth}
        \centering
        \includegraphics[width=\hsize]{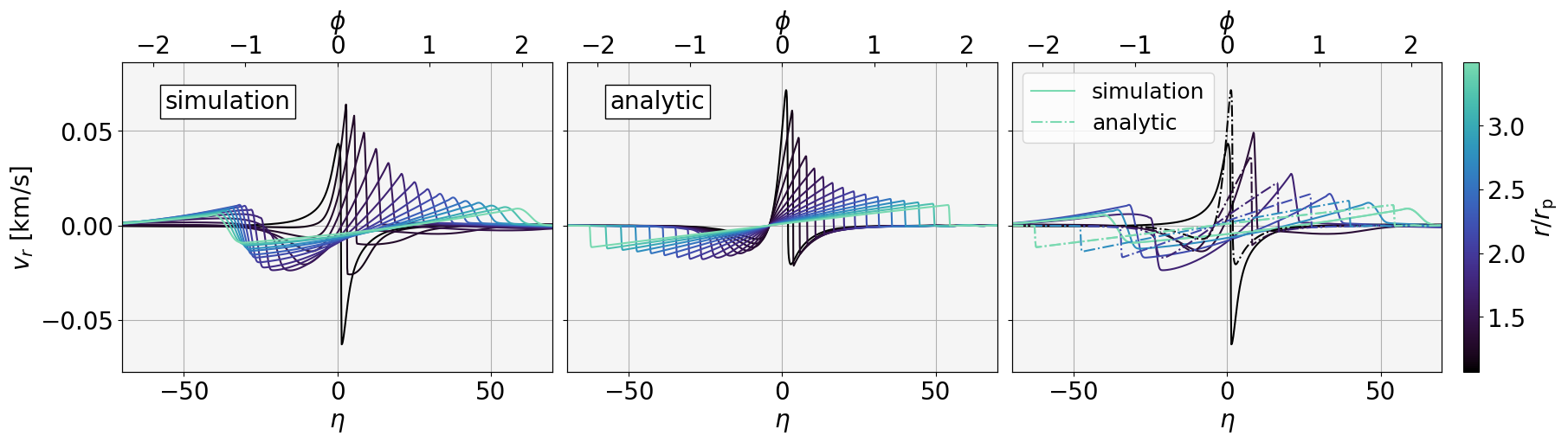}
        \caption{
              }
         \label{fig:Velocity Profiles 1.00}
         \end{subfigure}
        }
    \caption{Comparison between simulated (left) and analytical (middle) density (a) and radial velocity (b)  $\eta$ (azimuthal) profiles in the outer disc for the $M_{\rm p}=1.00\ m_{\rm th}$ case. In the right panels we show a choice of profiles from the simulation (solid line) and the analytic model (dash dotted line) to better highlight the differences in individual profiles. The color scale shows the radial distance from the planet. We filter the profiles from the simulation using the savgol\_filter function from the scipy Python library to reduce the oscillations at the shock fronts (see Appendix~\ref{App}).
              }
   \end{figure*}

   \begin{figure}
   \centering
   \includegraphics[width=\hsize]{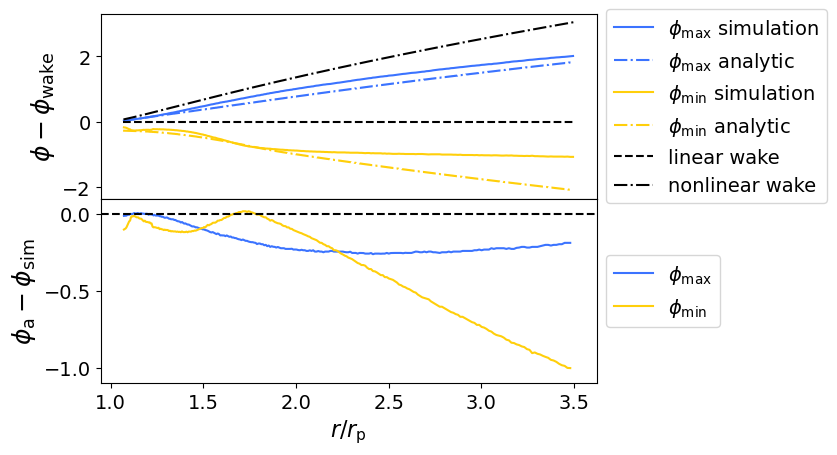}
      \caption{Top panel. Azimuthal distance of the maximum (blue) and minimum (yellow) from the linear wake prediction for the $M_{\rm p}=1.00\ m_{\rm th}$ case. We use this estimate to trace the pitch angle of the spiral wake. Solid lines represent the values obtained from the simulation, while dash dotted lines are obtained from the analytical model. Bottom panel: Difference between the maximum/minimum (blue/yellow) from the analytical model and the simulation as a function of the disc radius.
              }
         \label{fig:Pitch angle 1.00}
   \end{figure}

   We now consider the results we obtain when the planet mass approaches the thermal mass. The main assumption of the semi-analytical model we are testing is that the perturbations are linear close to the planet. This is equivalent to the requirement $m_{\rm p}<1~m_{\rm th}$. We test this limit to assess the validity of the model when the linear assumption is not satisfied.
   
   By comparing Figs.~\ref{fig:Density Map 1.00}-\ref{fig:Velocity Map 1.00} with Figs.~\ref{fig:Density Map 0.25}-\ref{fig:Velocity Map 0.25}, we see that the planet is starting to carve a gap in the disc. The density perturbations show positive features at the edge of the gap in the coorbital region, while the velocity shows material rotating around the planet position. In this limit, the spiral becomes wider and stronger, as expected for a more massive planet. Although the amplitude of the perturbations is now holding a better comparison between the simulation and the model, the discrepancy in pitch angle and width of the spiral increases. This is more evident in Figs.~\ref{fig:Density Profiles 1.00}-\ref{fig:Velocity Profiles 1.00}. The general features shown by the profiles are similar to the low mass case, with the exception of their width. In this case, the analytic profiles are wider than the simulation, where the center of the spiral suffers a stronger azimuthal shift. 
   
   In Fig.~\ref{fig:Pitch angle 1.00} we compare the azimuthal position of the minimum (yellow) and maximum (blue) of the density profiles with respect to the linear wake (black dashed line). We can see that the deviation between the simulation (solid lines) and the analytic model (dash-dotted lines) for the maximum increases up to $0.25$ radians and remains stationary after $2.0~ r/r_{\rm p}$. On the other hand, the deviation for the minimum has an oscillatory behaviour until $1.8~ r/r_{\rm p}$, after which it starts increasing until the outer edge of the disc, reaching a value of $1.00$ radians. This asymmetry not only changes the width of the profile, affecting the planet mass estimate\footnote{The azimuthal width of the profiles scales with the planet mass as $m_{\rm p}^{1/2}$ \citep{Bollati_2021}}, but it also results in a shift in pitch angle. The deviation of the pitch angle for spirals induced by massive planets from the linear prediction was also reported in the simulations of \citet{Zhu_2015,Bae_2018b}. We also plot the nonlinear correction from Eq.~\eqref{eq:Spiral Wake Nonlinear} (black dot dashed line) that was introduced in \citet{Cimerman_2021} to correct this behaviour. Although this nonlinear correction is valid for planet masses below the thermal limit, it deviates considerably from the peak of both the simulation and the analytical model above this limit. Thus it needs to be revisited in the limit of intermediate and high planet masses, $m_{\rm p}\geq m_{\rm th}$.

   %This implies that minimising the residuals between the simulation and the model won't be effective, as the perturbations will not overlap completely.

    \begin{figure}
    \centering
    \includegraphics[width=\hsize]{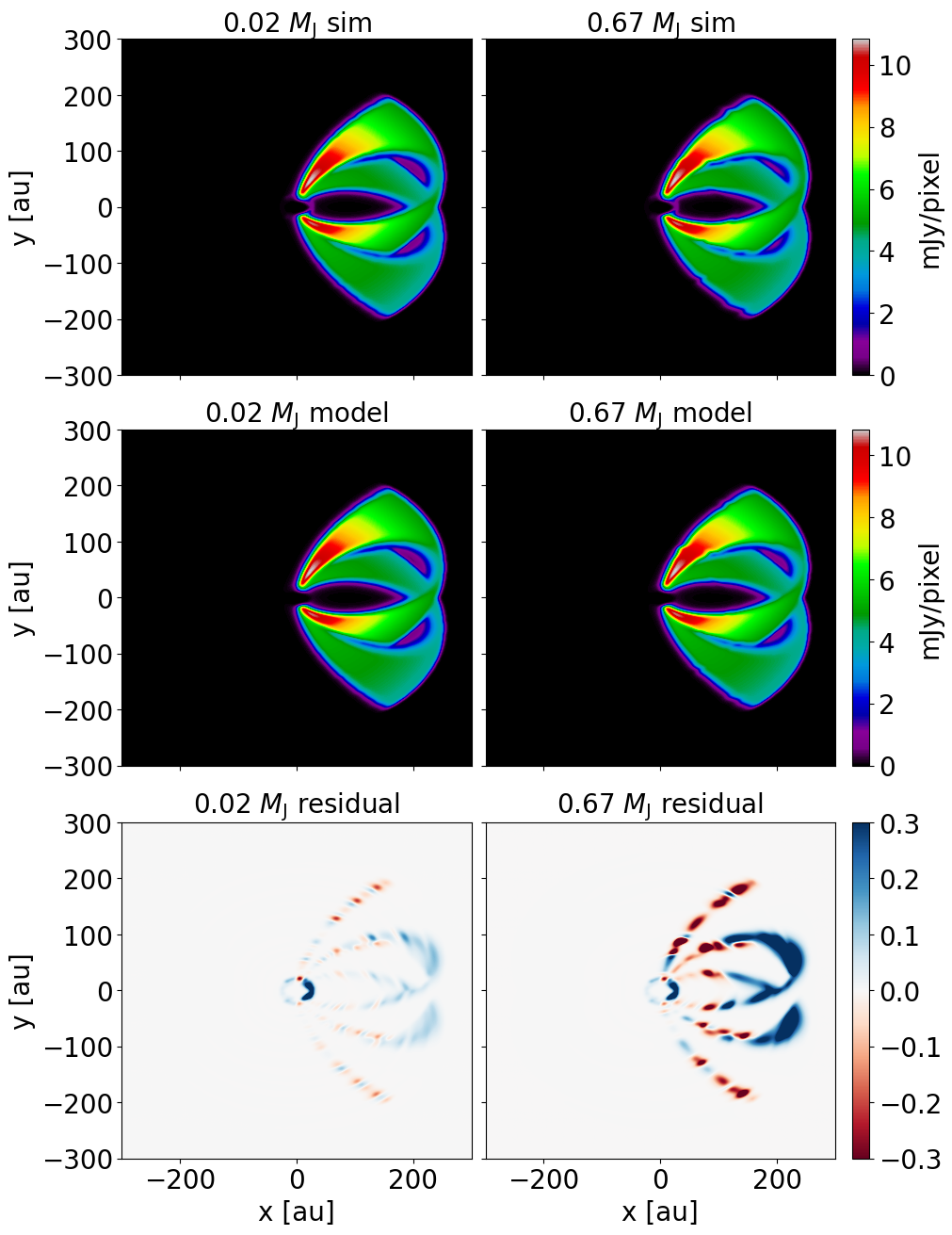}
        \caption{Expected kink in the channel maps for a thermal mass planet in a disc with $h_{\rm p}/r_{\rm p}=0.05$ (left column) and $h_{\rm p}/r_{\rm p}=0.1$ (right column), resulting in a planet mass of $m_{\rm p} \simeq 0.02~M_{\rm J}$ and $m_{\rm p} \simeq 0.67~M_{\rm J}$, respectively. We arbitrarily set the inclination of the disc to $i=-45\degree$. Top row: Channel maps obtained from the velocity field of the simulation. Middle row: Channel maps obtained from the velocity field of the analytical model. Bottom row: Intensity residuals between the channel maps obtained from the velocity field of the simulation and the analytic model.
              }
        \label{fig:Observability}
    \end{figure}
   
\subsection{Observational applications}
\label{sec:Wakeminer}

 Kinematic observations of protoplanetary discs have shown the presence of complex structures that can be associated to planet candidates \citep{Pinte_2020,Izquierdo_2023}. To visually test the applicability of the semi-analytical model to observations, we coupled \textsc{wakeflow} with \textsc{discminer} \citep{Izquierdo_2021}. The latter fits each channel map with a parametric model describing physical and geometrical properties of the disc. It is possible to use either a single or double Gaussian or Bell line profile model at each pixel, enabling to differentiate between the emission from the upper and lower surfaces of the disc. The \textsc{discminer} model assumes a Keplerian axisymmetric velocity field. In this way, it can minimise the residuals between the observations and the model in order to fit for the disc parameters, but it cannot predict planet properties such as the planet mass.
 
  We can model the presence of a planet by adding the velocity field computed with \textsc{wakeflow} on top of the background Keplerian field of \textsc{discminer}.  Once we have computed the perturbed field using our semi-analytical model, we can produce channel maps to perform a direct comparison with observed data. In a similar way, we can produce channel maps using the velocity field from the simulation.

  In Fig.~\ref{fig:Observability} we produce \textsc{discminer} channel maps using the simulation (top row) and \textsc{wakeflow} (middle row) velocity field. In the bottom row we also show the residuals between the two models. We show the output for a planet with $m_{\rm p} = 1~m_{\rm th}$ in a disc with $h_{\rm p}/r_{\rm p} = 0.05$ (left column) and $h_{\rm p}/r_{\rm p} = 0.1$ (right column), respectively. This is equivalent to having a planet with $m_{\rm p} \simeq 0.09~M_{\rm J}$ and $m_{\rm p} \simeq 0.7~M_{\rm J}$. We note how the perturbations in the first case have very low amplitude, thus are barely noticeable, compared to the disc with a more realistic disc aspect ratio. This is because the kink amplitude $\mathcal{A}$, defined as the maximum deviation from the unperturbed channel map, scales as $\mathcal{A}\propto m_{\rm p}^{1/2}$ \citep{Bollati_2021}, so the low planet mass excites perturbations with an amplitude of $\sim1\%$ of the Keplerian velocity at the planet location. This does not change the behaviour of the model in the different regimes compared to the thermal mass, but confirms how only planets in the Jupiter mass range or above can be detected with this method.  The  residuals in the bottom row highlight how the different pitch angle causes the spirals to cross the channel maps in different locations, producing oscillating residuals that do not cancel out when the amplitude of the spirals between the simulation and the model are comparable. As a result, a procedure that minimises this type of residuals cannot be used to estimate planet masses.

\subsection{Limitations of this work}
\label{sec:Limitations}

In the previous sections we have focused our analysis on the comparison between 2D hydrodynamical simulations with a globally isothermal equation of state and our analytic model, in order to test this framework using models sharing the same assumptions. However, real protoplanetary discs have a 3D structure and may feature cooling processes, likely resulting in vertical thermal stratification. Both the vertical structure of the disc \citep{Zhu_2015, Pinte_2019, Rosotti_2020, Rabago_2021} and a more realistic thermodynamics prescriptions \citep{Miranda_2019, Miranda_2020, Ziampras_2023} change the spiral structure generated by a planet, thus affecting planet mass estimates obtained from the analysis of its morphology. 

\citet{Muley_2024} have recently performed 3D hydrodynamical simulations comparing different thermodynamic prescriptions, from locally isothermal to $\beta$-cooling and three-temperature radiation hydrodynamics. They study the density and radial velocity azimuthal profiles at the midplane and at the emitting layer of $^{12}$CO for their different thermodynamic prescriptions and find that using more realistic equations of state produce perturbations in the midplane with lower amplitudes compared with the locally isothermal case. Above the midplane, instead, all the  prescriptions produce radial velocity perturbations with approximately the same amplitude, which is lower compared to the isothermal case in the midplane. As the planet mass is directly related to the amplitude of the perturbations \citep{Bollati_2021, Rabago_2021} and the analytical framework assumes that the emission comes from the midplane of a globally isothermal disc, then we expect that applying the analytical model to spirals coming from a finite height in a vertically temperature stratified disc will produce lower planet mass estimates.

Additionally, numerical 3D simulations have shown that planetary spirals are less tightly wound in the upper layers of the disc \citep{Zhu_2015,Rosotti_2020}, due to the vertical temperature gradient of the disc. Following our discussion in Sec.~\ref{sec:Wakeminer}, constraining the pitch angle dependence on the disc parameters also in the vertical direction is crucial to perform a fitting procedure using the analytical framework in order to obtain accurate estimates of planet masses.

%-----------------------------------------------------------------
\section{Conclusions}
\label{sec:Conclusions} 

The focus of the search for protoplanets is moving towards kinematic detections. With the development of new techniques to reveal protoplanet candidates, new possibilities open for the characterisation of their properties. Together with incoming observational constraints, novel and improved models need to be developed and robustly tested. In this paper we have numerically tested the semi-analytical model for planet induced spiral waves introduced in \citet{Bollati_2021} with a set of 2D \textsc{FARGO3D} simulations in the low ($<1~m_{\rm th}$) and intermediate ($\sim1~m_{\rm th}$) planet mass regime. In contrast to previous studies in the low planet mass regime \citep{Cimerman_2021}, our analysis aims to assess the applicability of this analytical model to observations by testing the velocity field in the intermediate mass regime, as this is the quantity directly detected by kinematic observations. The results we obtained can be summarised as follows:

\begin{itemize}
    \item The comparison with the simulation with a $0.25~m_{\rm th}$ planet shows similar results with those obtained in \citet{Cimerman_2021} for the density field. Although being qualitatively similar, the solution of Burgers' equation propagates faster with respect to the simulations, producing spirals with smaller width and amplitude. We find the same behaviour for the radial velocity component for a radial distance greater than $\sim 0.2~\rm r/r_{\rm p}$ from the planet.\\
    \item Both perturbations feature a positive azimuthal shift in the center of the spiral and a secondary spiral arm in the simulations, not present in the analytical model. This produces a change in pitch angle even for low planet masses.\\
    \item The velocity field computed from the semi-analytical model features a discontinuity at the interface between the linear and nonlinear regions. This discontinuity is present in both planetary regimes.\\
    \item For the simulation with a $1~m_{\rm th}$ planet, the discrepancy increases. The model produces profiles with smaller amplitude but larger width, and it fails to reproduce the shift in azimuth present in the simulation profiles. This results in a different pitch angle. The correction introduced in \citet{Cimerman_2021} does not reproduce the expected pitch angle from the simulation in this limit.
\end{itemize}

Although the general behaviour of the model is in good qualitative agreement with the simulation, the difference in pitch angle causes the two spirals not to overlap. As a result, standard fitting procedures based on the minimisation of residuals cannot converge. In order to apply this model to observations it is necessary to revisit it accounting for the shift in pitch angle and the discontinuity in the velocity field that are present for all planet masses. An alternative approach for fitting planet masses would consist in developing a parametric model for the velocity perturbations.

\begin{acknowledgements}
       We thank the referee for the constructive feedback provided that helped improve the manuscript. We also thank Jochen Stadler, Bin Ren and Enrico Ragusa for the useful discussions. This project has received funding from the European Research Council (ERC) under the European Union’s Horizon 2020 research and innovation programme (PROTOPLANETS, grant agreement No. 101002188). This work has received funding from the European Union’s Horizon 2020 research and innovation programme under the Marie Sklodowska-Curie grant agreement No 823823 (RISE DUSTBUSTERS project). AJW has received funding from the European Union’s Horizon 2020 research and innovation programme under the Marie Skłodowska-Curie grant agreement No 101104656. GR and AR acknowledge support from the European Union (ERC Starting Grant DiscEvol, project number 101039651) and from Fondazione Cariplo, grant No. 2022-1217. Views and opinions expressed are, however, those of the author(s) only and do not necessarily reflect those of the European Union or the European Research Council. Neither the European Union nor the granting authority can be held responsible for them. Computational resources have been provided by INDACO Core facility, which is a project of High Performance Computing at the Università degli Studi di Milano (https://www.unimi.it).

\end{acknowledgements}


%\begin{thebibliography}{}
\bibliographystyle{aa}
\bibliography{main}

%\end{thebibliography}

\begin{appendix} %First appendix
\section{High viscosity simulation}\label{App}
\begin{figure}
   \centering
       \begin{subfigure}[t]{\hsize}
       \centering
       \includegraphics[width=\hsize]{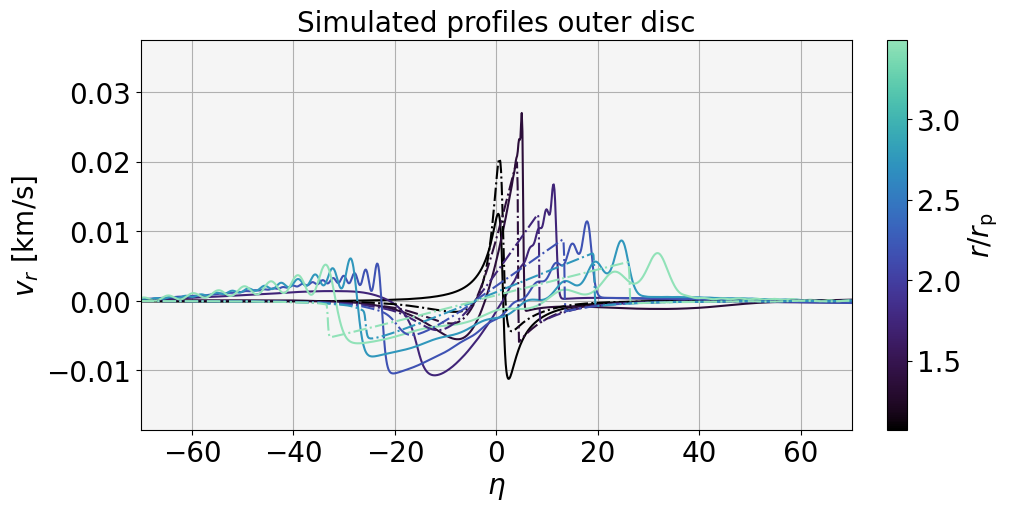}
          \caption{
                  }
             \label{fig:App_very_low_alpha}
        \end{subfigure}
   \hfill 
       \begin{subfigure}[t]{\hsize}
       \centering
       \includegraphics[width=\hsize]{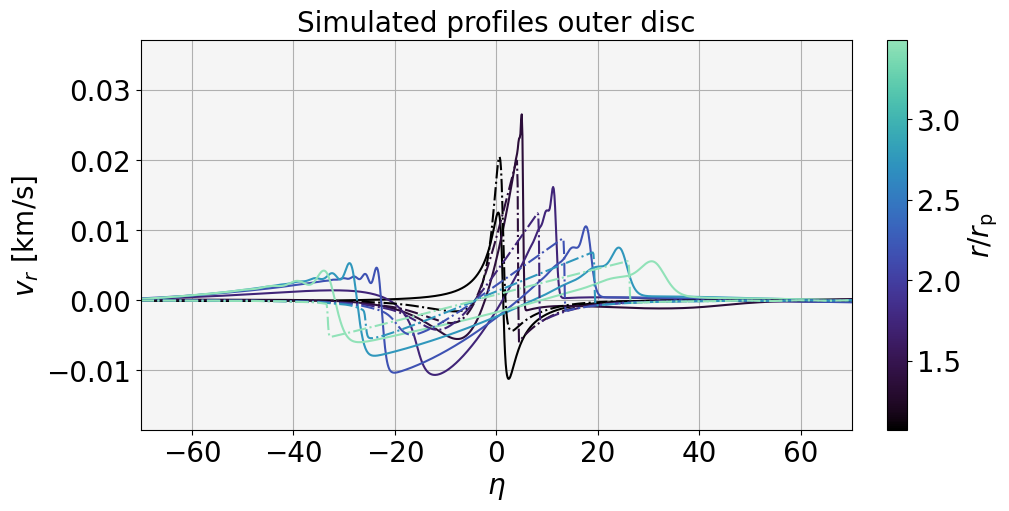}
          \caption{
                  }
             \label{fig:App_low_alpha}
        \end{subfigure}
   \hfill 
       \begin{subfigure}[t]{\hsize}
       \centering
       \includegraphics[width=\hsize]{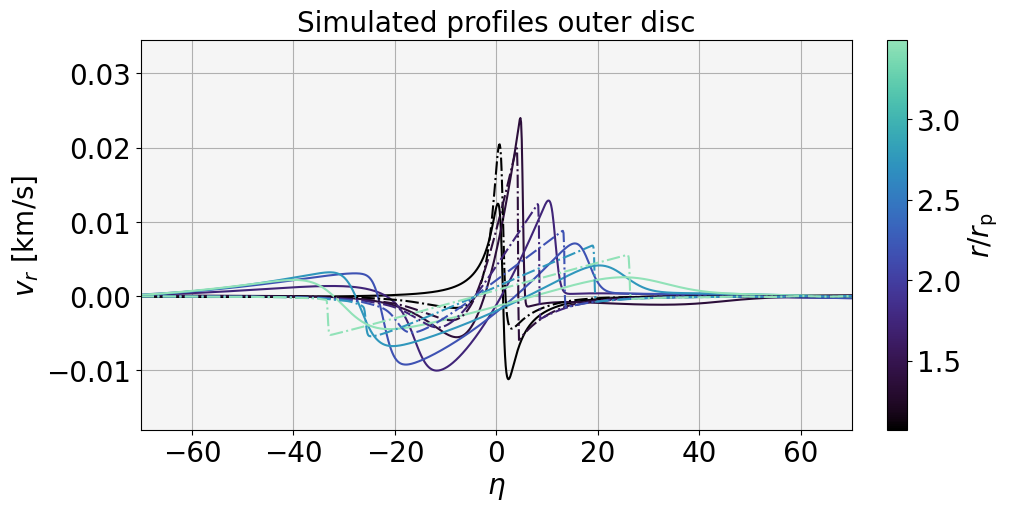}
          \caption{
                  }
             \label{fig:App_high_alpha}
        \end{subfigure}
    \caption{Comparison of a choice of radial velocity azimuthal profiles from the simulation (solid line) and the analytic model (dash dotted line) using $\alpha=0$ (a), $\alpha=10^{-4}$ (b) and $\alpha=10^{-3}$ (c).
                  }
   \end{figure}

In Fig.~\ref{fig:App_low_alpha} we show the third panel of Fig.~\ref{fig:Velocity Profiles 0.25} and compare it with the same result obtained using simulations with $\alpha=0$ and $\alpha=10^{-3}$ (Fig.~\ref{fig:App_high_alpha}). Here we show the full extent of the oscillations, whereas in the main text we filter them using the savgol\_filter function from the scipy Python library. While the profiles from the lower viscosity simulation is completely dominated by oscillations, the higher viscosity simulation does not show any sign of them. These oscillations are likely due to numerical instabilities and are damped out with increasing viscosity. However the viscous damping of the profiles becomes too strong for $\alpha=10^{-3}$, especially far away from the planet position, making a direct comparison between the analytical model and the simulations more challenging. As the structure of the perturbations is otherwise unchanged, we can safely carry out our analysis on the simulations with $\alpha=10^{-4}$.

\end{appendix}

\end{document}